\documentclass[aps]{revtex4}
\begin{document}
\def\ket#1{| #1\rangle}
\def\bra#1{\langle #1 |}
\def\Tr{{\rm{Tr}}}
\title{
Causal Quantum Theory and the Collapse Locality Loophole}
\author{ Adrian Kent}
\email{A.P.A.Kent@damtp.cam.ac.uk} 
\affiliation{Centre for Quantum Computation, Department of Applied 
Mathematics and Theoretical Physics, University of Cambridge,
Wilberforce Road, Cambridge, CB3 OWA, U.K.}
\altaffiliation[Affiliation when paper first drafted: \qquad] 
{Hewlett-Packard Laboratories, Filton Road, Stoke Gifford, 
Bristol, BS34 8QZ, U.K.}
\date{April 2002, revised March 2005}

\begin{abstract}
Causal quantum theory is an umbrella term for
ordinary quantum theory modified by two 
hypotheses: state vector reduction is a well-defined process, and
strict local causality applies.  The first of these holds
in some versions of Copenhagen quantum theory and need not necessarily
imply practically testable deviations from ordinary quantum theory.
The second implies that measurement events which are spacelike
separated have no non-local correlations.  To test this prediction, which 
sharply differs from standard quantum theory,
requires a precise definition of state vector reduction.

Formally speaking, any precise version of causal quantum 
theory defines a local hidden variable theory.
However, causal quantum theory is most naturally seen as a variant of standard
quantum theory.  For that reason it seems a more serious rival to 
standard quantum theory than local hidden variable 
models relying on the locality or detector efficiency loopholes.   

Some plausible versions of causal quantum theory
are not refuted by any Bell experiments to date, nor is it evident
that they are inconsistent with other experiments.  They evade
refutation via a neglected loophole in Bell experiments --- the {\it
collapse locality loophole} --- which exists because of the possible
time lag between a particle entering a measurement device and
a collapse taking place.  Fairly definitive tests of causal 
versus standard quantum theory could be made by observing 
entangled particles separated by $\approx 0.1$ light
seconds.
\end{abstract}
\maketitle
\section{Introduction}

The subtle relationship 
between quantum theory and relativity raises questions  
fundamental to our understanding of nature.  
Entanglement was first identified as a
potential source of tension between the two theories by
Einstein, Podolsky and Rosen \cite{epr}, while Bell's 
work \cite{bellone,belltwo} in the early
1960s made precise the sense in which classical intuitions
based on the principles of special relativity conflict with
quantum theory.  Theoretical and experimental investigations have
continued ever since.

After much careful analysis, a strong consensus emerged and has
held firm over the last two decades.
On this consensus view, insofar as special
relativity inspires us to consider alternatives to standard quantum
theory, those alternatives are characterised by Bell's definition of
local hidden variable theories.  However, the experimental evidence
very strongly favours quantum theory against local hidden variable
theories.  The hypothesis of local hidden variables can only be
maintained by supposing that a local hidden variable theory somehow
exploits one or more loopholes arising from our inability to construct
perfect experimental tests.  

Only two loopholes --- the detector efficiency \cite{detectionloophole} 
and locality loopholes
--- have generally been considered worth serious attention, and even
they are not generally thought to be plausible mechanisms for reconciling
local hidden variables with experiment.  Indeed, one 
recent experiment \cite{roweetal}
has succeeded in closing the detector efficiency loophole.  And,
although in principle the locality loophole can never be completely
closed, it has been substantially closed by another recent 
experiment \cite{weihsetal},
which implies that local hidden variable theories which use the
locality loophole would have to correlate the states of quantum random
number generators with those of the entangled particles being
measured.

Admittedly, no experiment to date has succeeded in simultaneously 
closing both loopholes, and there is a serious case for 
attempting still more stringent experiments (see 
e.g. \cite{aspect,vaidman} for discussions).  
Nonetheless, the general consensus 
is that a local hidden 
variable mechanism which exploits either or both loopholes
in a way which would not have shown up in experiments to date
would require a theory so perversely conspiratorial as to be 
almost incredible.

However, one or two gaps in this analysis have lately been noted. 
Altering standard causation, either by directly postulating reverse
causation \cite{price} or by considering statistically based
configuration space models \cite{akcracow}, allows alternatives to
local hidden variable theories that are consistent with relativity and
not excluded by Bell's theorem.  Also, a previously neglected loophole
in Bell experiments --- the memory loophole --- has lately been
identified \cite{accardiregoli,bchkp}.  However, no way of actually
reproducing quantum predictions within non-standard causation models
has been identified apart from ad hoc constructions \cite{akcracow}
that again appear perversely conspiratorial.  As for the memory
loophole, its potential effect, though real, is negligible when large
numbers of entangled particle pairs are tested \cite{bchkp,gill}.
Moreover, analysing the experimental data in a nonstandard but natural
way can eliminate the effect entirely \cite{bchkp,gill}.
Thus, though the memory loophole is an interesting subtlety, it does
not {\it per se} constitute a serious challenge to the 
standard interpretation of Bell experiments.  Its late
discovery should, though, at least disturb the general confidence that
absolutely everything was sorted out by Bell's and Clauser et al.'s
analyses \cite{bellone,belltwo,chsh} and subsequent experiments (for
example \cite{aspectetal,titteletal,weihsetal}).

Non-standard causation models, too, have at least one virtue: they
illustrate that considering new physical principles can suggest new
ways of thinking about non-local correlations.  One can too easily fall
into the habit of caricaturing local hidden variable theories as
involving small classical particles flying from source to measuring
device, carrying tables of instructions telling them what to do when
measured.  For theories that exploit the locality loophole, the
caricature version has little signalling devices sitting in the
experimental apparatus, sending signals to something like a radio
receiver attached to the particles, to inform them of prematurely made
random choices.  When the detector efficiency loophole is used, the
caricature version equips the particles with probes which identify the
detector, calculate its efficiency, and adjust the instructions tables
accordingly.  These pictures are indeed fantastically conspiratorial.
But nothing in the mathematical analysis of non-local correlations
implies that hidden variables theories have to work like this.  
Despite their admitted defects, proposals like reverse causation \cite{price}
or statistical configuration models defined by local 
weightings \cite{akcracow} do at
least illustrate the possibility of a different sort of story.

\section{Causal quantum theory and the collapse locality loophole} 

This paper considers another gap in the analysis of entanglement and
non-local correlations --- one which seems more serious 
than any of the loopholes previously considered.  
This is the possibility that state reduction is a well-defined
physical process, localised in space-time, and that, once this 
definition is taken into account, strict local causality (in Bell's 
sense \cite{bellbeables,bellfreewill}) holds.  
{\it Causal quantum
theory} is a useful umbrella term for the class of theories that arise
in this way, modulo various possible definitions of state reduction. 

The strict local causality hypothesis implies in particular that,
if $P_{\rm causal~qt}  ( A | \Lambda_P ; \psi ( - \infty ) )$ is the
probability of a state reduction $A$ taking 
place at a point $P$ in space-time, given all the state
reduction events in the past light cone $\Lambda_P$ of $P$ and
the initial state at $t = - \infty$, and if $B$ is any collection
of state reduction events taking place at points spacelike
separated from $P$, then 
\begin{equation} 
P_{\rm causal~qt}  ( A | \Lambda_P ; \psi ( - \infty ) ) = 
P_{\rm causal~qt}  ( A | \Lambda_P ; \psi ( - \infty ) ; B ) \, . 
\end{equation} 
In other words, and contrary to standard quantum theory, 
state reduction involves no non-local correlations.   
On the other hand, local state reduction probabilities themselves,
conditioned on past light cone events, should agree with those 
predicted by standard quantum theory, after perhaps allowing for
some modification (which we will assume to be slight in the 
cases we consider here) arising from introducing a precise
definition of state reduction:  
\begin{equation} 
P_{\rm causal~qt} ( A | \Lambda_P ; \psi ( - \infty ) ) \approx
P_{\rm standard~qt} ( A | \Lambda_P ; \psi ( - \infty ) ) \, . 
\end{equation} 
 
Consider for example two widely separated particles prepared in
a state close to (but, for reasons which will become apparent
later, not precisely equal to) a singlet: 
\begin{equation}
\ket{\psi} \approx 
\frac{1}{\sqrt{2}} ( \ket{0}_A \ket{1}_B - \ket{1}_A \ket{0}_B ) \, .
\end{equation}
As we will discuss in more detail below, according to causal quantum theory, if the two particles are 
measured in the $\ket{0}, \ket{1}$ basis, and {\it if the state
reductions corresponding to the measurements are completed at 
space-like separations}, the joint outcome probabilities 
are 
\begin{equation}
P_{\rm causal~qt}  ( 0_A , 0_B ) \approx
P_{\rm causal~qt}  ( 0_A , 1_B ) \approx
P_{\rm causal~qt}  ( 1_A , 0_B ) \approx
P_{\rm causal~qt}  ( 1_A , 1_B ) \approx 1/4 \, , 
\end{equation} 
whereas 
\begin{equation}
P_{\rm standard~qt}  ( 0_A , 1_B ) \approx 
P_{\rm standard~qt}  ( 1_A , 0_B ) \approx 1/2 \, , \qquad  
P_{\rm standard~qt}  ( 0_A , 0_B ) \approx 
P_{\rm standard~qt}  ( 1_A , 1_B ) \approx 0 \, . 
\end{equation} 
It seems that this gross discrepancy should show up
immediately in experiments on entangled particles: there is 
no need even to vary the measurement choices at $A$ and $B$ in
order to see the difference between the two theories.  
Given the impressive confirmation of quantum theory in Bell 
experiments to date, what is the point in considering causal quantum
theory?  

The loophole is in the italicized qualification: 
{\it if the state
reductions corresponding to the measurements are completed at 
space-like separations}.  
We can't be sure that state reductions occur at 
space-like separations, without knowing precisely when
they take place --- in other words, without a precise theory of 
state reduction.  And if the state reduction at
$A$ in fact takes place in the past light cone of that at $B$,
or vice versa, then  
\begin{equation}
P_{\rm causal~qt}  ( 0_A , 1_B ) \approx 
P_{\rm causal~qt}  ( 1_A , 0_B ) \approx 1/2 \, , \qquad 
P_{\rm causal~qt}  ( 0_A , 0_B ) \approx 
P_{\rm causal~qt}  ( 1_A , 1_B ) \approx 0 \, . 
\end{equation} 
in agreement with standard quantum theory.  More generally, 
causal quantum theory will agree with standard quantum theory 
in any Bell experiment, so long as the state reductions
for particles $A$ and $B$ in any given pair are timelike separated.  

It is worth emphasizing that the loophole causal quantum theory exploits 
is quite distinct from the locality loophole, mentioned above.  
The locality loophole relies on the fact that it is 
difficult to arrange a Bell experiment so that the measurements 
at $A$ and $B$ are chosen randomly and independently for each
pair, in such a way that the random choices are themselves 
made at points which are space-like separated from one another
and from the point at which the measured singlet is created. 
As in principle one could imagine that a common source far in the
past correlates hidden variables determining the particles' 
actions with those determining the outcomes produced by 
any randomiser, it seems impossible to close the locality 
loophole completely.   However, one can aim to close it beyond
any reasonable doubt, and much experimental ingenuity has been devoted
to doing so, from the famous
experiments of Aspect et al. \cite{aspectetal} utilising the 
quasi-randomness of high frequency waves, to recent experiments
using fast quantum random number generators \cite{weihsetal} or
passive quantum switches \cite{titteletal}.

The loophole considered here also
involves locality, but it involves the problem of ensuring that 
the state reduction {\it events} associated with measurements 
are spacelike separated, rather than ensuring that
randomly made measurement {\it choices} are.  
Let us call it the {\it collapse locality loophole}. 

\section{Is causal quantum theory self-consistent?} 

The basic features of causal quantum theory are best illustrated
in the idealised model of quantum states and measurements, commonly
used in quantum information theory, in which subsystems are 
treated as effectively pointlike and measurements are carried
out at a definite point in space and time.  To simplify the notation
further, we can also assume that the subsystems are stationary 
relative to one another, and that the hamiltonian is zero.  
Since the predictions of causal quantum theory depend on 
past events, we need to specify a state history as well as
a state.  We could assume that there exists at some time $t=0$ a state 
which has been undisturbed, and has not spontaneously collapsed,
for long enough that causal and standard quantum theory initially agree. 
This may be a realistic assumption for the state of an isolated
microscopic subsystem.  In principle, though, the fundamentally
correct formulation of causal quantum theory requires as an initial
condition the specification of the quantum state of the entire
system on some initial spacelike hypersurface.   

Suppose, in any case, that at $t=0$ 
we have a system in a state
\begin{equation}
\ket{\Psi (0) } = \sum_{i_1 \ldots i_n} a_{i_1 \ldots i_n} \ket{i_1}_1
\ldots \ket{i_n}_n \, , 
\end{equation}
where $1, \ldots , n$ denote distinct fixed points $x_1 , \ldots , x_n$  
in space and $i_1 , \ldots , i_n$ label internal degrees of freedom.
Note that we have chosen here a system whose internal degrees of freedom
may be entangled but whose spatial wave functions are unentangled:  
this is not essential for our discussion, but simplifies the
notation. 
Suppose too that the system has been undisturbed --- in
particular, no measurement event has taken place --- for a time
long compared to any of the subsystem separations. 

In principle, causal quantum theory can be defined for any theory
of state vector reduction in which the reductions are localised. 
However, to simplify the following discussion, we will suppose that the
only relevant state reductions here correspond to measurements of the internal
degrees of freedom, and we will assume that the subsystems remain
in essentially the same locations throughout: in particular, we will ignore
the evolution and eventual spreading of their spatial wave
functions.   We thus now suppose that we have some definite theory of 
state reduction which tells us precisely when a measurement takes 
place on any subsystem, and characterises the nature of the 
measurement, and that the theory tells us that no other reductions
are relevant for the system in question.   

The theory's prescriptions take the following form: a state reduction 
takes place at $(\underline{x_i} , t_i )$, defined by a set of 
operators $\{ A_{j} \}$ 
which obey  
\begin{equation}
\sum_j  ( A_{j} )^{\dagger} A_{j} = I \, , 
\end{equation} 
and which act on the Hilbert space corresponding to the internal
degrees of freedom of particle $i$.   
We define  ${\cal A}_{ij}$ to be the corresponding operator on the 
tensor product Hilbert space: that is, 
\begin{equation} 
{\cal A}_{ij} = I \otimes \ldots \otimes I \otimes A_{j} 
              \otimes I \otimes \ldots \otimes I \, , 
\end{equation}
where $A_{j}$ is the $i$-th term in the product.  

In standard quantum theory, whenever such a reduction takes place on
a state $\ket{\Psi}$, we 
get outcome $j$ with probability 
\begin{equation}\label{prob}
\Tr ( ( {\cal A}_{ij} )^{\dagger} {\cal A}_{ij} \rho_{\Psi} ) \, ,
\end{equation} 
where $\rho_{\Psi} = \ket{\Psi} \bra{\Psi}$.  
After this reduction, the state becomes
\begin{equation}\label{reduction}
\frac{ {\cal A}_{ij} \ket{\Psi} }{ ( \Tr ( ( {\cal A}_{ij} )^{\dagger} 
{\cal A}_{ij} \rho_{\Psi} )
)^{1/2} } \, . 
\end{equation} 
In between collapses, since the hamiltonian is zero, 
the state remains constant. 

In the corresponding version of causal quantum theory, in order to calculate
outcome probabilities for a measurement taking place
at $P_i = ( x_i , t_i )$, we first need to calculate a causally
defined version of the standard quantum state --- let us call it  
the {\it local state} --- of the system at $P_i$.  This is obtained
by starting from $\ket{\Psi (0) }$ and then applying (\ref{reduction})
for each measurement within the past light cone of $P_i$, sequentially
in time order, but no 
others: by assumption, the reductions are localised, so that
(assuming, of course, that we have a Lorentz invariant reduction theory) the
result is independent of the choice of frame which defines the 
time ordering.  Equation (\ref{prob}), applied to the local state, then defines
the outcome probabilities for the measurement at $P_i$.   Note that 
we do {\it not} define these probabilities in terms of the evolved
state vector $\ket{\Psi (t_i )}$, as we would in standard quantum theory.  

To calculate the joint probabilities for outcomes of measurements 
at mutually space-like separated points
$P_{i_1} , \ldots , P_{i_n}$, we need first to calculate the 
probability of each possible configuration of measurements,
and each possible set of outcomes, in the union of the past
light cones of the $P_i$: let us refer to these collectively
as ${\rm past~data}$.  We can calculate the conditional
probability of outcome $O_{i_j}$ at point $P_{i_j}$ given
the past data, $P ( O_{i_j} \, | \, {\rm past~data} \, )$. 
The joint probability of outcomes $O_{i_1} , \ldots , O_{i_n}$ is 
\begin{equation}
\prod_{j=1}^n P ( O_{i_j} \, | \, {\rm past~data} \, ) \, .  
\end{equation}  

Using these rules (and in principle using some small meshing
of space-time, and taking the limit as the mesh size tends
to zero), we can (in principle) proceed iteratively to calculate
the probabilities of all possible configurations of measurement
events in space-time. 

But does this procedure always produce well-defined answers?  
Suppose for example that we
begin with two subsystems in a singlet state  
\begin{equation} 
\ket{\Psi (0)} =  
\frac{1}{\sqrt{2}} ( \ket{0}_1 \ket{1}_2 - \ket{1}_1 \ket{0}_2 ) \, .
\end{equation}
Suppose no reductions take place before time $t$, and
that at time $t$, reductions take place at {\it both} 
$x_1$ and $x_2$ and that in both cases the reduction operators 
are $\{ P_0 , P_1 \}$, the projections onto $\ket{0}$ and
$\ket{1}$.  One possibility, which according to our rules has probability $1/4$, 
is that the measurement outcome in both cases will be that
corresponding to the operator $P_0$.  If no further 
reductions take place before time $T= t+ | x_1 - x_2 |$, then
at that point our rules suggest the local states at the
points $(x_i , T)$ both become 
\begin{equation}
\frac{P_0 \otimes P_0 \ket{\Psi (0)}}{{ | P_0 \otimes P_0 \ket{\Psi (0)} | }}
 \, , 
\end{equation}  
which is undefined.  After this point, we thus have no
rule for predicting future measurement outcomes.  
A similar problem arises if both outcomes correspond to $P_1$, of
course.   

There are two attitudes one can take to this.   One is to conclude
that causal quantum theory is not a properly defined theory, and 
deserves no further attention.  The other is to note that, in practice
and even in principle, the theory can be saved quite easily.  

A practical counter-argument to the above follows from the 
fact that the singlet state is never precisely realised in nature.  
A more realistic version of the above discussion would thus begin
with  
$$
\ket{\Psi (0) } = \sum_{ij} a_{ij} \ket{i}_1 \ket{j}_2 \, , 
$$ 
where $a_{01} \approx 1/\sqrt{2} \approx a_{10}$ and
$a_{00} \approx 0 \approx a_{11}$, but neither of the 
last two terms are precisely zero.  A still more realistic
version would take the initial state to be a mixture, dominated
by states of approximately this form.   Either way, the local states after
measurement are, by virtue of the correction terms, well-defined.  

More generally, whatever measurement operators arise in a theory 
of reduction, one would not ever expect to find in nature --- or 
to be able to create artificially ---  a state that is 
{\it precisely} a zero eigenstate of a tensor product of 
non-zero operators.  
 
An arguably more principled way of avoiding the difficulty is to require
that the theory of state reduction should involve only
measurement operators which have no zero eigenvalues.  (So, in particular,
it cannot include projections.)  The Ghirardi-Rimini-Weber
spontaneous localisation model \cite{grw} is an example of such a theory. 
(Note, though, that in current non-relativistic versions of the GRW 
model, the measurement operators are not perfectly localised, so
that the probability of a collapse event centred at a point $P$ 
is only approximately determined by events in the past light cone of $P$: it depends
also to some extent on events some way outside the past light cone.)    
In such theories, although the cumulative effect of 
measurements can asympotically tend to the action of a projection,
no collection of measurement events ever completely annihilates 
the component of the state in any given subspace.  
In particular, whatever the initial state, the calculations
in the causal version of such theories will never produce 
a zero value for the unnormalised local state, and 
the local states are always defined. 

Put succinctly, the worry about causal quantum theory was that 
it might imply combinations of measurement outcomes that are 
impossible in standard quantum theory --- and that when that
happens, causal quantum theory breaks down.  The way around
this is to notice that in practice the 
measurement outcomes which occur in causal quantum theory
will almost surely never actually be impossible in 
standard quantum theory: if one prefers, by slightly restricting
the theory of reduction one can ensure this is always true.  
However, unless the details of the 
reduction theory somehow prevent long-range entanglement, 
combinations of outcomes which are 
extremely unlikely according to standard theory can be expected to be 
fairly common in causal quantum theory.  

\section{Testing causal quantum theory}

As noted above, causal quantum theory and standard quantum theory
predict different outcome probabilities for separated measurements on
entangled states, so that Bell experiments seem the first obvious place
to look for a refutation of causal quantum theory.  

The greatest
separation over which apparently non-local correlations have so far
been demonstrated was in the beautifully designed experiments of Tittel et
al. \cite{titteletal} and Zbinden et al. \cite{zbindenetal}, who 
have demonstrated the violation of Bell
inequalities, and the confirmation of quantum predictions, by
entangled photons separated by $ \approx 10$ km, or $ \approx 3 \times
10^{-5}$ light seconds.  Tittel et al.'s experimental arrangement was
somewhat asymmetric.   Zbinden et al.'s experiment was 
designed to ensure that the photons arrive at separated 
detectors within a time interval of less than $5$ ps (in 
laboratory frame): however, the relevant detectors here were surfaces
coated in absorbing black paint rather than standard photo-detectors.   
This experiment was designed to test an intuition inspired by
ideas of Suarez and Scarani \cite{suarezscarani}.  Zbinden 
et al.'s experiment tests and refutes the hypothesis that 
correlations different from those of quantum theory arise
when each absorbing paint detector carries out the measurement first,
from the perspective of its own stationary reference frame.

As Zbinden et al. carefully note, some admittedly questionable,
although plausible, assumptions are needed in order to justify
interpreting their experiment as equivalent to a standard Bell
experiment with photo-detectors.  
Let us in any case make the best case assumption that the 
results of Tittel et al. and Zbinden et al. imply that a
standard Bell experiment could be carried out with effectively identical apparatus on
both sides, such that entangled photons on both sides enter
standard photo-detectors, or in an alternative version of the
experiment hit patches of 
absorbing paint that function as detectors, at precisely the same times, and that the results would still confirm
standard quantum theory.  If we could be sure that any sensible theory
of state reduction implies that reduction takes place within the
photo-detectors, or alternatively within the absorbing paint, 
in a time shorter than $3 \times 10^{-5}$ seconds --- or even merely that the times at
which it takes place in the two detectors are separated by $< 3 \times
10^{-5}$ seconds --- then we could conclude from these results that
causal quantum theory was definitely refuted.

But can we be so sure?  All ideas about theories of state reduction 
are speculative, but among them, at least three have been taken
seriously from time to time by a significant number of thoughtful 
people: Wigner's 
suggestion \cite{wigner} 
that state reduction could be somehow caused by conscious minds, Penrose's
suggestion \cite{penrose} that state reduction takes place when required to
prevent a superposition of macroscopically distinct gravitational 
fields, and Ghirardi-Rimini-Weber-Pearle type theories in which state
reduction results from a spontaneous localisation process 
occurring stochastically at rates proportional to particle number or mass.  

Insofar as these
ideas can be made precise at all, none of them seems necessarily to imply    
state reductions in photo-detectors that are necessarily separated
by times short compared to $3 \times 10^{-5}$ sec.  
Indeed, if Wigner's
suggestion were right, reduction wouldn't occur at 
all until experimenters look at the
data.  The other two cases cannot properly be analysed without
a detailed description of the apparatus.  However, given that 
the reduction of a superposition in the GRWP and Penrose theories 
depends on the extent to which it involves macroscopically distinct 
separations of massive particles in position space, it would be
surprising if very tight bounds on the reduction time in these
experiments could be derived.  Even in an experiment with perfect
symmetry between the two wings, in which the photons enter 
the photo-detectors at the same time $t$ in the experimental 
rest frame, it need not necessarily be the case that the collapse
events also take place at the same time $t + \delta$ --- the time
$\delta$ before collapse could, as in GRWP theories, be stochastically
determined, with independent stochastic processes associated to
space-like separated points on the two wings.  
Using absorbing black paint in place of a photo-detector almost
certainly would only make the numerics worse, since a photon hitting an 
absorbing surface does not at all quickly create a macroscopically
distinct configuration of massive particles in position space.    

Still, if one takes the idea of a state reduction theory
seriously in the first place then, whatever one thinks of  
Wigner's suggestion, there is a good reason to assume that reduction
{\it does} take place (at the very latest) not long after the
impression of a measurement result registers in a human observer's
brain --- namely, our own experience.  When we watch an 
apparatus carrying out measurements, 
it seems to us as though each measurement produces a definite result, 
and it seems to us that these results are accessible to us rather
soon after the point at which the signal reaches our eye or ear.  

Of course, this does not logically imply that a state reduction
has taken place.  It could conceivably be that we enter a superposition state,
entangled with the apparatus and the measured system, at least for
some time, but that the properties of our consciousness are such that
it constructs for us the impression of quickly accessible definite
results before reduction takes place.  But once one entertains {\it this}
hypothesis, there seems no reason to postulate state reduction at
all.  One might then as well go all the way, and follow 
Everettians in assuming that there is
only unitary evolution, but that the properties of consciousness
are such that we perceive things according to one component
of the universal state vector, in which definite measurement results 
took place and were observed by us.  (See e.g. Ref. \cite{wallace} and
references therein for recent discussions advocating this view.) 

On this reasoning, any state reduction 
theory worth taking seriously should imply that reduction
ordinarily would take place within $\approx 0.1$ sec --- roughly the 
timescale over which we can discriminate events --- of the signal
from a measurement apparatus reaching us.   Given this, a fairly
definitive test of causal quantum theory could be carried out
by allowing observers separated by $\approx 0.1$ light seconds
to carry out synchronized measurements on entangled particles 
and directly observe the results, before later comparing them.    

\section{Conclusions}

The standard case for studying loopholes in Bell experiments is 
that quantum non-locality has such fundamental significance that
it is worth demonstrating as rigorously as possible.  Even highly
implausible alternative explanations are worth analysing and, if
possible, refuting.  

A more practical motivation has also recently been
suggested \cite{mayersyao,gisinprivcomm}.  It may be crucial for future
users of quantum cryptography and quantum communication systems to
guard against fakery or sabotage by testing that states involving
allegedly entangled separated subsystems genuinely are entangled
states of the correct form.  In principle, Bell experiments can do
this.  But, again in principle, a saboteur might make use of any Bell
experiment loophole to produce apparent, but unreliable, evidence of
entanglement.  

While these are certainly sufficient motivations for considering the
collapse locality loophole, I think there are stronger reasons.  For there
is a principled case for taking seriously both the hypotheses which
define causal quantum theory.  

Take first the idea that there is an explicit physical theory of state
reduction.  This is not, by any means, everyone's preferred solution
to the measurement problem --- but it is a natural solution which has
often been advocated.  Indeed, almost everyone who has studied the
Copenhagen interpretation must have wondered whether one could not
replace the projection postulate, with its vague reference to
measurement, with a precise physical law.  Granted, devising explicit
collapse models is a project fraught with difficulties.  It seems hard
to find satisfactorily relativistic versions of GRWP models (see e.g.
Ref. \cite{pearlequasirel} for a recent review).  It is also hard to
precisely formulate Penrose's idea, let alone Wigner's.  These are
very serious worries.  But so far {\it every} proposed solution to the
measurement problem is fraught with difficulties, and yet presumably
there must be a solution.
  
As for the idea that strict local causality should hold: this is,
obviously, inspired by special relativity.  Of course, we have learned
that quantum theory respects Minkowski causality more subtly.  But
strict local causality remains a natural hypothesis --- albeit, unless
nature has indeed deceived us by exploiting the collapse locality
loophole, an incorrect one. 

Against causal quantum theory, it might be argued there is something
decidedly strange about a theory which --- a harsh critic could say
--- maintains consistency only by relying on the existence of small
errors or by its own inability ever to give completely definite
answers.  Maybe: but then many features of standard quantum theory
seem strange until, perhaps, familiarity breeds acceptance.
Experimental evidence would be more convincing than arguments
based on aesthetic preconceptions, particularly as the aesthetic
arguments are not entirely one-sided.    

A stronger argument, perhaps, is that, if causal quantum theory
were right, one might expect detectable consequences other than
in Bell-like experiments on entangled states.   For instance, particles whose
wave functions have extended support ought, so to speak, sometimes
to appear to collapse and localise in two or more places at once.  
For the reasons already
discussed, this need not lead to logical contradiction.  For example,
a causal quantum theory analysis could suggest that 
an apparent observation of a particular particle most likely derived from some 
other source, or that it was most likely due to detector error, 
even when these explanations would be very unlikely
according to standard quantum theory.  But one might
think it ought to have observable consequences, in cosmology 
and elsewhere, which might be or have already been contradicted
by experiment.  Very possibly it does: the question deserves
careful analysis.  

Another fair argument is that it seems that, for causal
quantum theory to be right, and yet not to have been detected in
Bell experiments to date, the relevant parameters in the hypothetical
explicit collapse model would have to be relatively fine-tuned.  
Bell experiments with separations of $> 10^{-5}$ light
seconds have been performed, with no detectable deviation from
standard quantum theory; an experiment with separation of $\approx 10^{-1}$
light seconds should show dramatic deviations from quantum theory
if causal quantum theory were correct.   It would be a bit of a 
quirk of fate for the critical separation to lie in a range
covered by fewer than four orders of magnitude.   
Of course, the type
of measurement carried out in the experiments is crucial.
The argument for the latter experiment sufficing relies on 
direct observation of the results by human observers, rather 
than the photo-detectors used in the former.  
According to Wigner's suggestion, this makes all the difference, and 
so the fine-tuning argument does not apply against some Wignerian version
of causal quantum theory. 
But, from a GRWP or Penrosean collapse model 
perspective, it isn't obvious that the human brain should be hugely 
better at inducing collapse than a photo-detector, and so the
fine-tuning argument has some force in these cases.  

At the moment, though, these counterarguments don't seem totally
compelling.  Although the arguments against local hidden variable
theories exploiting the detector efficiency and (standard) locality
loopholes are stronger, experimentalists rightly continue to work
towards more definitive tests.  The Earth is large enough to allow
almost, if not absolutely, conclusive experimental tests of causal
versus standard quantum theory; if (at least) one part of the experiment were
carried out on a short manned space flight, a completely definitive
test could be made.  
It would be fascinating to see the experiments done, even if they do
no more than remove a sliver of doubt.  One hopes they will be, once
technology allows suitably long range distribution of entanglement.  
\vskip10pt
\leftline{{\bf Acknowledgments}} \qquad This work was supported by a
Royal Society University Research Fellowships, 
projects EQUIP and PROSECCO (IST-2001-39227) of the IST-FET 
programme of the EC, and the Cambridge-MIT Institute.   
I thank Nicolas Gisin and Philip Pearle 
for helpful discussions, and the
Perimeter Institute for support while revising this paper.

\end{document}